\documentclass[letterpaper,twocolumn,10pt]{article}

\usepackage[hyphens]{url}
\usepackage{usenix,epsfig,hyperref,caption,subcaption}
\hypersetup{
    bookmarksnumbered=true,
    unicode=false,          
    pdftoolbar=true,        
    pdfmenubar=true,        
    pdffitwindow=false,     
    pdfstartview={FitH},    
    pdftitle={A Story of Suo Motos, Judicial Activism, and Article 184 (3)},   
    pdfnewwindow=true,      
    colorlinks=false,       
    linkcolor=red,          
    citecolor=green,        
    filecolor=magenta,      
    urlcolor=cyan           
}
\begin{document}
%
\title{\Large \bf A Story of Suo Motos, Judicial Activism, and Article 184 (3)\\
\small{An Analysis of a Decade of Open Judgements from the
Supreme Court of Pakistan}}

\author{
{\rm Zubair Nabi}\\
zubair.nabi@cantab.net
}


\date{}

\maketitle

\begin{abstract}
The synergy between Big Data and Open Data has the potential to revolutionize
information access in the developing world. Following this mantra, we present
the analysis of more than a decade worth of open judgements and orders from the
Supreme Court of Pakistan. Our overarching goal is to discern the presence of
judicial activism in the country in the wake of the \emph{Lawyers' Movement}.
Using Apache Spark as the processing engine we analyze hundreds of unstructured
PDF documents to sketch the evolution and various organs of judicial activism in
Pakistan since 2009. Our results show that the judiciary has indeed been
pursuing matters of public interest, especially those that pertain to the
fundamental rights of the citizens.
Furthermore, we show how the size of the presiding bench in a case and citations
of Articles from the Constitution and prior judgements can aid in classifying
legal judgements. Throughout the analysis we also highlight the challenges that
anyone who aims to apply Big Data techniques to Open Data will face. We hope
that this work will be one in a series of community efforts to use Open Data as
a lens to analyze real-world events and phenomena in the developing world.
\end{abstract}


\section{Introduction}
Big Data analysis has been a boon to a large number of scientific domains,
ranging from genome sequencing and particle physics~\cite{vivien:2013:btb} to
transportation management~\cite{Biem:2010:IIS} and urban
planning~\cite{RePEc:pio:envirb}. Originally spawned by web 2.0 companies, such
as social networks, e-commerce sites, and search engines, to cater to their
business needs, Big Data has now pervaded almost every academic, scientific, and
business field. This ``Big Data Spring'' has been aided by an entire ecosystem
of largely open source tools, including processing
frameworks~\cite{Dean:2004:MSD,Zaharia:2012:RDD}, storage
solutions~\cite{Lakshman:2010:CDS, hive}, and analytical
tools~\cite{Olston:2008:PLN,Ihaka:1996:R}. A large number of recent
initiatives~\cite{Berlingerio:2013:AAA,Eagle:2014:BDF} have made use of Big Data
analytics to tackle problems in the developing world. While these endeavours are
indeed promising, the lack of open data, human and technical capacity, and
finances in the developing world has incapacitated it from truly riding the Big
Data wave~\cite{VWC:2012:BDB,Piotrowski:2014:BOA}.


The primary problem is the lack of data in the first place. In most of the
developing world, census data for instance is manually collected via surveys,
which are rarely digitized~\cite{UNESC:2014:BDA}. Furthermore, even if the data
is collected and digitized, it is locked away behind multiple levels of
bureaucratic red tape. Similarly, private sector entities, such as telcos have a
rich set of data, even in these regions but privacy and unclear data sharing
incentives have hampered their ``data philanthropy''
efforts~\cite{VWC:2012:BDB}. Finally, if the data is available, the lack of
analytics and systems skills in the local workforce and almost non-existent data
management policy at the state level, hinders any useful analysis. The problem
is exacerbated by traditional technology-centric developing world problems, such
as low Internet penetration and intermittent power, which fortify the digital
divide in data representation~\cite{Piotrowski:2014:BOA}.

In the last few years, aided by technology, a number of developing world
countries have experienced popular uprisings centered around the devolution of
power. In certain cases, the judiciary has also played an active role in the
quest to counter and even depose decades old authoritarian regimes and political
structures. For instance, even as early as 2006, before the ``Arab Spring'', the
judiciary in Egypt started questioning the status quo. Dubbed the \emph{Judicial
Intifada}, members of the Judges Club clashed with the government of Hosni
Mubarak for the independence and reform of the judiciary~\cite{Eldin:2006:AJI}.
Similarly, the 2007 \emph{Lawyers' Movement} in Pakistan also aimed to establish
the independence of the judiciary in a country where the courts have
traditionally towed the line of the government. In the wake of the movement, the
superior judiciary started confronting the government of the time in matters of
public interest. Similar instances of judicial activism or public interest
litigation are applicable to India, Malaysia, and Turkey, among many
others~\cite{Khan:2010:SB}.

The legal field in general has also embraced Big Data by making use of it to,
for instance, predict the outcome of cases and motions, and determine litigation
fees~\cite{Dysart:2013:HLA}. Some experts have gone to the extent of calling
predictive Big Data analytics the ``potential holy grail in the practice of
law''~\cite{Nelson:2013:BDB}. Unfortunately, so far only private firms have been
able to leverage these solutions whereas widespread public sector usage has been
found wanting due to the lack of open data and tools. This has resulted in
public practitioners and society at large becoming legal Big Data
``poor''~\cite{boyd:2012:CQF}. More concretely, the lack of a ``Legal Big Data
Toolkit'' to incorporate legal complexities, jargon, and nuance is a major
challenge~\cite{Sheridan:2014:BDF}. 

In this paper, we tackle this dual Big Data ``poverty''--in the developing world
and the legal domain--by analyzing legal judgements from the Supreme Court of
Pakistan--the apex court of the country--which span more than a decade of open
data. In the process, we highlight (1) the challenges faced in analyzing
unstructured and noisy PDF documents, (2) the use of open source Big Data tools
for developing world problems, and (3) the need to possess domain specific
knowledge, for instance to interpret legal jargon, while undertaking such an
analysis. While only a preliminary study, our results show that even limited
open data can reveal a wide range of interesting trends and figures.
Specifically, our analysis shows that the number of Suo Moto cases--wherein the
court decides to take up a case of its own accord--increased after a popular
judicial revival in the country. In addition, the bulk of the judgements pertain
to constitutional, civil, and human rights cases, indicating a tilt towards
public interest issues. The same trend is reflected by the Articles of the
Constitution and previous judgements frequently referred to in the judgements.
Finally, we also report trends in the bench distribution, jurisdiction,
judgements per year, prolific judges, and others.

The rest of the paper is organized as follows. In the next section
(\S\ref{sec:background}), we give a brief introduction to the judicial system in
Pakistan. In addition, we provide an overview of Apache
Spark~\cite{Zaharia:2012:RDD}, the data processing platform we employed for our
analysis. The methodology of the study is presented in \S\ref{sec:method}.
\S\ref{sec:meta} unveils the analysis of the metadata from the documents. These
results are augmented by the in-depth analysis of the actual content in
\S\ref{sec:indepth}. \S\ref{sec:conclusion} concludes the paper and outlines
future work.

\section{Background}\label{sec:background}
In the section, we give a high-level overview of the legal system in Pakistan
(\S\ref{subsec:judicial}). We follow this up with an introduction to Apache
Spark, an all-encompassing Big Data processing platform (\S\ref{subsec:spark}).

\subsection{Judicial System of Pakistan}\label{subsec:judicial}
The Judicial System of Pakistan is divided into the Superior Judiciary,
Subordinate Courts, and Special Courts and Tribunals~\cite{Hussain:2011:TJS}. As
this paper analyzes judgements from the Supreme Court, we discuss its set up in
detail while glossing over the details of the Subordinate Courts, and Special
Courts and Tribunals. The Superior Judiciary consists of the Supreme Court, 5
High Courts (one for each of the 4 provinces and the Federal District), and the
Federal Shariat Court. This set up is comprehensively laid out and enshrined in
the Constitution of Pakistan which also enunciates  the ``separation of
judiciary from the executive'' and the ``independence of
judiciary''~\cite{Hussain:2011:TJS}. The Superior Judiciary also enjoys
financial autonomy from the Executive and the Legislative branches of
government, allowing it to act independently. Appointments to the Superior
Judiciary are  handled by an independent Judicial Commission--comprising senior
judges and representatives from the bar and government--which makes
recommendations which are then confirmed by the Legislature. Similarly,
accountability is enforced through a Supreme Judicial Council which is made up
of senior judges from the Superior Judiciary.

The Supreme Court, which is the apex court of the country, has original,
appellate, and advisory jurisdiction. It has original jurisdiction in
inter-governmental disputes and fundamental rights issues, appellate
jurisdiction in civil and criminal issues, and advisory jurisdiction in legal
and constitutional advice to the Government. Its decisions and interpretations
are final and binding on all other courts. The Superior Court comprises a Chief
Justice and a bench of 16 judges. Similar structure, with some differences,
applies to the various High Courts. The Federal Shariat Court on the other hand,
only has jurisdiction to decide ``whether or not a certain provision of law is
repugnant to the injunctions of Islam''~\cite{Hussain:2011:TJS}. The Subordinate
Courts deal with civil and criminal matters. Finally, a number of Special Courts
and Tribunals, such as Banking Courts, Income Tax Appellate Tribunal, and
Anti-Terrorism Courts have jurisdiction over domain specific issues and matters.

The Constitution of Pakistan has elements from British Law--largely through the
Government of India Act 1935--and Islamic Law~\cite{khan2005constitutional}.
Post-Independence in 1947, the Government of India Act 1935 was amended and
Islamic Provisions were incorporated over a period of many decades to result in
the 1973 Constitution. This Constitution has undergone 20 Amendments since then
to reach its current state. It consists of 280 Articles, divided into
Sections, which collectively enshrine the setup and structure of the state and
system, and the rights of the citizens\footnote{The Constitution of the Islamic
Republic of Pakistan -- \url{http://www.pakistani.org/pakistan/constitution/}}.

Over the course of its turbulent history, Pakistan has oscillated between
democratic and military rule. During every military rule, the judiciary has been
accused of not only turning a blind eye to the military regime but even
providing legal cover to it under the ``Doctrine of State
Necessity"~\cite{Kumar:2007:JSH}. All of this dramatically changed in 2007, when
the Chief Justice of the Supreme Court, Iftikhar Muhammad Chaudhry, refused to
resign on the orders of the then President, General Pervez Musharraf. Without
getting into the specific details of the incident and its aftermath, it will
suffice to say that the subsequent sacking of the Superior Judiciary, spawned a
popular movement, led by lawyers and students, which culminated in the
restoration of all judges including the Chief Justice in 2009~\cite{TPL}.
Cognizant of their public mandate, the bench started pursuing issues of public
interest through Suo Moto cases. Again, without getting into the discussion of
the pros and cons of this judicial activism~\cite{Economist:2011:PPJ}, in this
paper, we use the judgements from the last one decade as a lens to illuminate
its various facets. Therefore, the goal of this paper is neither to argue for a
stance on judicial activism nor to question its ramifications or to take a
position in any of the events.

\begin{figure*}[t]
\centering
  \includegraphics[width=1\linewidth]{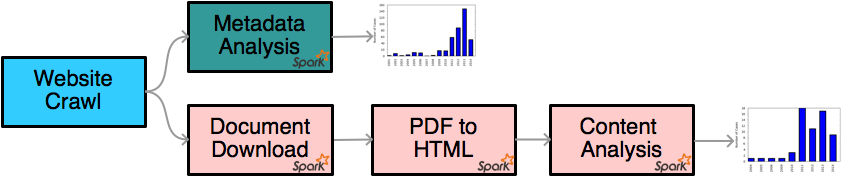}
  \caption{Processing Pipeline}
  \label{fig:processing}
\end{figure*}

\subsection{Apache Spark}\label{subsec:spark}
Apache Spark is an open source data processing platform, originally spawned at
UC Berkeley~\cite{Zaharia:2012:RDD}, which incorporates multiple processing
paradigms, including batch, iterative, and streaming analysis. Like its
predecessor, Hadoop\footnote{Apache Hadoop -- \url{http://hadoop.apache.org/}},
it masks away the intricacies of distributed computation, such as work distribution, scheduling, data transfer, and
fault-tolerance, beneath a simple API. In essence, the application developer
only focuses on the business logic of the application, while the underlying
platform performs all the heavy lifting. Unlike Hadoop, which has a strict
two-stage API (\emph{map} and then \emph{reduce}, with a transparent \emph{group
by} in the middle), Spark applications can have multiple stages, thus forming a
directed graph of computation. At the very core of Spark, lies the concept of
\emph{Resilient Distributed Datasets} (RDDs), which are datasets with
customizable persistence, fault-tolerance, and availability guarantees.
RDDs can be cached in memory (with selective compression and serialization) or
backed by flat files on a distributed filesystem (\emph{Hadoop Distributed
Filesystem} (HDFS)~\cite{Shvachko:2010:HDF} by default). Each RDD also stores
its lineage graph, allowing it to be regenerated if it is lost. This ability to
process data in memory allows Spark, in some cases, to achieve an order of
magnitude better performance than Hadoop~\cite{Zaharia:2012:RDD}.

A Spark application starts by converting a dataset, typically read from the
HDFS, to an RDD. A functional API is then used to perform transformations on
this dataset, where each transformation results in a new RDD which is ingested
by a subsequent stage. Each transformation (or the application of a function) is
parallelized across multiple cores and machines. The computation is exhausted
when a ``side-effect'', writing to a file or printing to standard output for
instance, is introduced into the computation graph. The core of Spark is
implemented in Scala~\cite{odersky2004overview}, a functional programming
language. Scala is also the language of choice for implementing Spark
applications, although it has bindings for other popular languages, such as Java
and Python. In the last few years, the core Spark project has evolved to add
stream processing~\cite{Zaharia:2013:DSF}, declarative processing and
storage~\cite{Xin:2013:SSR}, and a suite of machine learning\footnote{Spark
MLlib -- \url{https://spark.apache.org/mllib/}} and
graph processing\footnote{Spark GraphX --
\url{https://spark.apache.org/graphx/}} libraries, to its standard arsenal.

\section{Methodology}\label{sec:method}
In this section, we discuss the methodology we adopted for our analysis
including data collection, curation, and examination. 

Since 2001, the Supreme Court of Pakistan has been hosting some of its
judgements and orders on its
website\footnote{http://www.supremecourt.gov.pk/web/page.asp?id=103}. These
documents are available in the form of unstructured PDFs often augmented with a
small amount of metadata. The metadata consists of the title of the
judgement/order, its public release date, and a brief description. We followed a
five step analysis: (1) website crawl, (2) metadata analysis, (3) document
download, (4) conversion from PDF to HTML, and (5) analysis of the content of
the HMTL documents. In the following, we briefly discuss each step.
\subsection{Website Crawl}
All of the judgements and orders are linked from a single page. This page
contains the metadata for each document as well as a link to a separate download
page for each. We scraped the HTML from this page using a custom Scala script to
generate a CSV file with four columns: (1) Download page link, (2) title, (3)
date, and (4) description (if available). In total we were able to get metadata
for 415 documents, starting from April 2001 and ending in August 2014.
\subsection{Metadata Analysis}
A Spark application was used to analyze this CSV to calculate the number of
cases per year, the distribution of Suo Moto cases, and the type of each
judgement/order (constitution, civil, criminal, etc.)
\subsection{Document Download}
We then wrote another Spark job to download each PDF and store it locally. 12
out of the 415 links were dead, reducing the number of downloaded PDFs to 403.
\subsection{Conversion of PDF to HTML}
Apache PDFBox\footnote{Apache PDFBox -- \url{https://pdfbox.apache.org/}} was
employed for the PDF to HTML conversion. A Spark job was applied to convert the 403 PDFs to HTML and then store them locally
again. Out of these, 25 documents could not be converted as they were in
Urdu--the national language of Pakistan. In addition, the conversion for another
7 PDFs failed due to bad quality of their PDF images. Therefore, a total of 371
HTML documents were obtained.
\subsection{Content Analysis}
A Document Object Model\footnote{W3C Document Object Model (DOM) --
\url{http://www.w3.org/DOM/}} for each HTML document was generated and used in
conjunction with a series of regular expressions to extract relevant information
through a series of Spark jobs. Using this procedure, the jurisdiction of each
case, the distribution of the panel for each case, and the Articles of the
Constitution and previous judgements/orders referred to in each document were
computed.
\subsection{Discussion}
We initially implemented each of these stages separately and used Spark for
exploratory and interactive analysis of the data. Once we had all the building
blocks in place, we connected them together to engender an automated processing
pipeline (depicted in Figure~\ref{fig:processing}). Therefore, the same
application can be used in an automated fashion to analyze other similar
datasets.

At this point, it is necessary to highlight that 403 PDF documents can by no
means be labelled as ``Big Data'' and the use of a data intensive computing
platform, such as Spark might be overkill~\cite{Rowstron:2012:NEG}. But the
general techniques and tools used in our analysis can be used to analyze a
larger number of such documents. For instance, each of the 5 High Courts also
has its judgements/orders on its website. Collectively, this dataset comprises
thousands of documents. In addition, the same analysis can be extended to India,
which has a similar judicial system. In case of India, judgements for both the
Supreme Court as well as dozens of its High Courts since 1950 are available as
open data, potentially resulting in a dataset with millions of such
documents\footnote{The Judgments Information System of India --
\url{http://judis.nic.in/}}. Although in this paper, we confine ourselves to the
said 403 documents, we hope that our thesis, proposed practices, and lessons
will serve as a blueprint for others to undertake similar analysis. Along with
the legal field, these techniques can be augmented to analyze open data in the
developing world from a wider range of domains, including agriculture, energy,
education, and health-care. Such studies will be extremely useful in
illustrating various trends and statistics which can be put to the test in the
real-world to improve some of the shortcomings in these domains. Open data is
extremely promising, but unless it is made consumable, it is meaningless.

A similar argument applies to the usage of Spark. We could easily have written a
standalone Scala application to perform the analysis. We opted for a standard
distributed processing platform to underscore how such platforms can simplify
application development and potentially crunch a similar but substantially
larger dataset.

\begin{figure}[t]
\centering
  \includegraphics[width=1.0\linewidth]{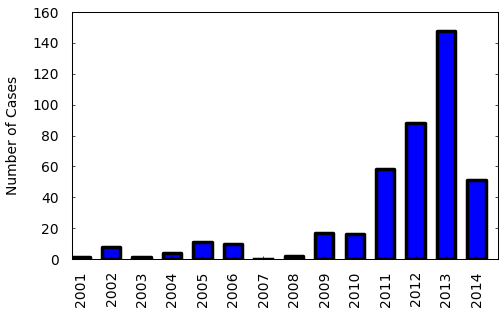}
  \caption{Distribution of cases by year.}
  \label{fig:year}
\end{figure}

\section{Metadata Analysis}\label{sec:meta}
This section presents the outcome of our metadata analysis.

\subsection{Distribution of cases by year}
As mentioned before, the metadata also contains a public release date for the
judgements. While this date was present in most instances, it was found to be
missing in a few. We manually filled in the missing values by cross-referencing
them with other online sources. This leads to the first lesson that we learnt:
\emph{L1 -- Expect missing values in open data}. Figure~\ref{fig:year} plots the
distribution of cases by year. We see a sharp increase from 2009 onwards in the
number of judgements/orders passed annually. This coincides with the
restoration of the judges in 2009. Due to its wide jurisdiction, the Supreme
Court of Pakistan has traditionally had to deal with more cases than its
capacity. It annually receives more than 15000 petitions and appeals, and more
than 30000 applications~\cite{Hussain:2011:TJS}. Most of these are routed to
provincial benches of the Court to ensure fast response.
Only important cases are dealt with by the Principal Bench--whose judgements
constitute our dataset. Therefore, the increase in the number of
judgements/orders post-restoration in 2009, indicates the inclination of the
Supreme Court to take up more cases in the wider public interest as a form of
judicial activism. It may also be indicative of the effectiveness of the
National Judicial Policy enacted in 2009 to expedite and streamline the Superior
Judicial Process~\cite{Hussain:2011:TJS}.

\begin{figure}[t]
\centering
  \includegraphics[width=1.0\linewidth]{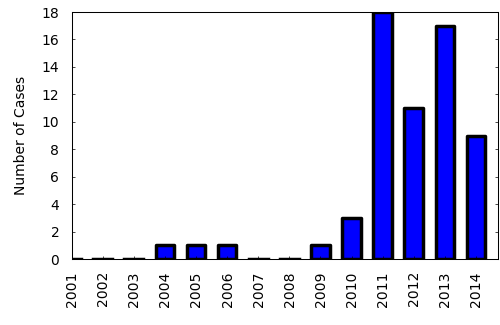}
  \caption{Distribution of Suo Moto cases by year.}
  \label{fig:yearsmc}
\end{figure}

One way to discern judicial activism is by analyzing the number of Suo Moto
cases undertaken by the Court. Under Article 184 (3) of the Constitution, the
Supreme Court has the power to take up cases of its own accord which violate
the fundamental rights of the people as enunciated in the Constitution.
Typically, these are undertaken after a story appears in the print and/or electronic media
or a citizen directly writes to the Court~\cite{suomoto}. Ordinarily it is the
job of the State through its policing system to ensure the fundamental rights
of its citizens. But if it is ineffective in doing so, the void can be filled by
the judiciary to play a more active role in the resolution of such issues.
Figure~\ref{fig:yearsmc}, segments the distribution of Suo Moto cases by year.
In line with Figure~\ref{fig:year}, the number of Suo Moto cases also follows a
similar trend, which suggests the presence of judicial activism. In addition,
Suo Moto cases accounted for 8.1\% of the total cases pre-2009. Post-2009, once
the deposed bench was reinstated, that number increased to 15.6\%, indicating a
clear tendency of the court to exercise Article 184 (3).

\begin{table}[t]
\centering
    \begin{tabular}{| l | l | l |}
    \hline
    \textbf{S\#} & \textbf{Type} & \textbf{\# of Cases} \\
	\hline
	1. & Constitution & 173 \\
	2. & Civil Miscellaneous Application & 99 \\
	3. & Suo Moto Case & 62 \\
	4. & Human Rights Case & 48 \\
	5. & Civil & 41 \\
	6. & Civil Appeal & 39 \\
	7. & Civil Review & 16 \\
	8. & Criminal & 12 \\
	9. & Criminal Appeal & 11 \\
	10. & Criminal Miscellaneous Application & 9 \\
	11. & Reference & 6 \\
	12. & Unknown & 3 \\
	13. & Jail Petition & 2 \\
	14. & Civil Petition for Leave to Appeal & 1 \\
    \hline
    \end{tabular}
    \caption{Distribution of cases by type.}
    \label{tab:casetype}
\end{table}

\subsection{Distribution of cases by type}
The title of each case in the metadata also contains its type in many instances.
In total we were able to glean 13 different types. It is noteworthy that (a) a
case might be made up of more than one type; for instance a case might be both a
Civil Miscellaneous Application as well as a Constitution Petition, and (b) the
title only contains the abbreviations of a few of these types from within a
case. For instance, we noticed 12 different descriptions for Constitution
Petition, including a number of misspelt variants. In addition, there is
inconsistency in the syntax used to describe a case. Furthermore, some of the
types were abbreviated. In some instances, it was easy to expand these
abbreviations. For example, \emph{Const. Petition} is clearly the abbreviated
form of Constitution Petition. Consider another example, \emph{c.p.} which may
map to Civil Petition, Constitution Petition, or Criminal Petition. To ensure
that all variants were accounted for, we constructed a look-up table for this
resolution by consulting various online sources, including the Supreme Court
website itself. This exercise resulted in three lessons: \emph{L2 -- There will
be term inconsistency in open data}, \emph{L3 -- Terms will often be misspelt},
and \emph{L4 -- Domain knowledge--legal jargon in this case--is priceless in
even basic analysis}. All in all, metadata is not enough or is an inaccurate
measure to gauge all of the different types that constitute the
judgements/orders.
Nonetheless, it provides an initial window into the distribution of these
different types. These are presented in Table~\ref{tab:casetype}. Constitution
Petitions, for which the Court has original jurisdiction, make up around
1/3rd of all judgements/orders. In addition, Suo Moto and Human Rights cases
also have a sizable presence, mostly at the discretion of the Court, indicating
its interest in pursuing issues concerning fundamental rights.

\section{In-depth Analysis}\label{sec:indepth}
To perform an in-depth analysis of the actual content of the documents, we first
converted them from PDF to HTML. As noted earlier, out of total 403, 371 were
successfully converted to HTML. 32 documents were skipped primarily because they
were in Urdu, even though this was not mentioned in the metadata: \emph{L5 --
Open data contains incomplete and even misleading metadata information}. 

\begin{table}[t]
\centering
    \begin{tabular}{| l | l | l |}
    \hline
    \textbf{S\#} & \textbf{Jurisdiction} & \textbf{\# of Cases} \\
	\hline
	1. & Original & 230 \\
	2. & Appellate & 103 \\
	3. & Review & 18 \\
	4. & Original/Appellate & 9 \\
	5. & Advisory & 5 \\
	6. & Original/Review & 4 \\
	7. & Contempt & 1 \\
	8. & Unknown & 1 \\
    \hline
    \end{tabular}
    \caption{Distribution of cases by jurisdiction.}
    \label{tab:casejurisdiction}
\end{table} 

\subsection{Distribution by jurisdiction}
Each judgement/order also contains the jurisdiction of the case. This
information is useful in analyzing the different number of cases for which the
Court had direct jurisdiction, i.e. constitutional and fundamental rights
issues, or indirect jurisdiction, in case of civil and criminal matters. The
distribution of cases by jurisdiction gleaned from the content of the documents
is shown in Table~\ref{tab:casejurisdiction}. This presents another view of the
data in Table~\ref{tab:casetype}. The number of cases for which the court has
original jurisdiction--constitution, Suo Moto, and human rights--make up the
bulk of the judgements similar to the distribution by type.

\begin{table}[t]
\centering
    \begin{tabular}{| l | l | l |}
    \hline
    \textbf{S\#} & \textbf{Name} & \textbf{\# of Cases} \\
	\hline
	1. & Justice Iftikhar Muhammad Chaudhry & 232 \\
	2. & Justice Jawwad S. Khawaja & 125 \\
	3. & Justice Khilji Arif Hussain & 95 \\
	4. & Justice Gulzar Ahmed & 75 \\
	5. & Justice Tassaduq Hussain Jillani & 74 \\
	6. & Justice Sh. Azmat Saeed & 71 \\
	7. & Justice Ijaz Ahmed Chaudhry & 51 \\
	8. & Justice Hani Muslim & 50 \\
	9. & Justice Tariq Parvez & 50 \\
	10. & Justice Anwar Zaheer Jamali & 45 \\
    \hline
    \end{tabular}
    \caption{List of the ten most prolific judges.}
    \label{tab:bench}
\end{table} 

\subsection{Most prolific judges}
We next analyze the distribution of the make up of the bench for each case. The
PDF documents also contain this information. The DOM for each HTML document was
not sufficient for extracting this information due to its extreme, unstructured
nature: \emph{L6 -- Open data can potentially be unstructured and hard to
parse}. Therefore, we had to augment the DOM with a series of regular
expressions to extract the names of the presiding judges from the preamble of
each document. We could only extract these names from 364 out of the 371
documents; they were missing from 2 documents, while 5 documents had extremely
dirty, noisy formatting. In total we noted 52 different judges who collectively
presided over these cases.
The average size of the bench was 3 while the maximum was 17--the full bench.
The full bench was constituted in 10 instances which highlights the importance
of these cases. A quick inspection of these cases revealed that they were indeed
high profile. One of the cases dealt with the National Reconciliation Order (NRO), a
controversial law enacted by President Pervez Musharraf to provide carpet
amnesty to all politicians and bureaucrats accused of corruption and other
unlawful activities.
Similarly, another set of cases revolved around the 18th Amendment. The 18th
Amendment was applied to the Constitution in 2010 and among many other changes,
it revoked the power of the President to dissolve the legislature, and did away
with most of the changes introduced to the Constitution during the time of
President Pervez Musharraf. Both of these cases are clearly in the interest of
the wider public; thus illustrating that even the size of the bench can reflect
the importance of a case.

To work out the most prolific judges in the last one decade, we counted the
number of times they presided over a case in the overall dataset. Some of the
names were misspelt or abbreviated; thus underscoring the ubiquity of \emph{L2} and
\emph{L3}. Table~\ref{tab:bench} enumerates the ten most prolific judges.
The Chief Justice, Justice Iftikhar Muhammad Chaudhry, who was deposed by
President Pervez Musharraf, is by far the prolific judge by having presided over
almost half of all cases.

\begin{table}[t]
\centering
    \begin{tabular}{| l | l | l | l |}
    \hline
    \textbf{S\#} & \textbf{Article \#} & \textbf{CategorY
    (Chapter)} &
    \textbf{\# of Cases}
    \\
	\hline
	1. & 184 (3) & The Supreme Court & 107 \\
	2. & 199 & The High Courts & 62 \\
	3. & 9 & Fundamental Rights & 60 \\
	4. & 4 & Introductory & 44 \\
	5. & 184 & The Supreme Court  & 40 \\
	6. & 25 & Fundamental Rights & 38 \\
	7. & 3 & Introductory & 33 \\
	8. & 2A & Introductory & 29 \\
	9. & 14 & Fundamental Rights & 28 \\
	10. & 9 & Fundamental Rights & 26 \\
    \hline
    \end{tabular}
    \caption{List of the ten most cited Articles of the Constitution.}
    \label{tab:articles}
\end{table}

\begin{table*}[t]
\centering
    \begin{tabular}{| l | l | p{5cm} | l |}
    \hline
    \textbf{S\#} & \textbf{Article of Constitution} & \textbf{Description} &
    \textbf{\# of Cases}
    \\
	\hline
	1. & PLD 2009 SC 879 & Sindh High Court Bar Association vs Federation of
	Pakistan & 7 \\
	2. & PLD 2011 SC 997 & Watan Party vs Federation of Pakistan & 7 \\
	3. & PLD 1972 SC 139 & Asma Jilani vs The Government of Punjab & 6 \\
	4. & PLD 1996 SC 324 & Al-Jehad Trust vs Federation of Pakistan & 6 \\
	5. & PLD 1988 SC 416 & Benazir Bhutto vs Federation of Pakistan & 5 \\
	6. & PLD 1973 SC 236 & Raunaq Ali vs Chief Settlement Commissioner & 4 \\
	7. & PLD 1979 SC 38 & Zulfiqar Ali Bhutto vs The State & 3 \\
	8. & PLD 1983 SC 457 & Fauji Foundation vs Shamimur Rehman & 3 \\
	9. & PLD 1993 SC 210 & KBCA vs Hashwani Sales and Services & 3 \\
	10. & PLD 2007 SC 578 & Chief Justice of Pakistan vs President of Pakistan &
	3 \\
    \hline
    \end{tabular}
    \caption{List of the ten most cited PLD judgements.}
    \label{tab:pld}
\end{table*}

\begin{table*}[t]
\centering
    \begin{tabular}{| l | l | p{5cm} | l |}
    \hline
    \textbf{S\#} & \textbf{Article of Constitution} & \textbf{Description} &
    \textbf{\# of Cases}
    \\
	\hline
	1. & 1991 SCMR 1041 & I.A. Sherwani vs Government of Pakistan & 14 \\
	2. & 1998 SCMR 793 & Syed Zulfiqar Mehdi vs Pakistan International
	Airlines Corporation & 12 \\
	3. & 2010 SCMR 1301 & Tariq Aziz ud Din vs Federation of Pakistan & 11 \\
	4. & 1998 SCMR 2268 & Airport Support Services vs The Airport Manager & 7 \\
	5. & 2012 SCMR 773 & Alleged Corruption in Rental Power Plants & 7 \\
	6. & 1992 SCMR 563 & Inam-ur-Rehman vs Federation of Pakistan & 6 \\
	7. & 1999 SCMR 2883 & Adreshir Cowasjee vs KBCA & 6 \\
	8. & 1994 SCMR 1299 & Ghulam Mustafa Jatoi vs Additional District and Sessions
	Judge & 5 \\
	9. & 1997 SCMR 641 & Gadoon Textile Mills vs WAPDA & 5 \\
	10. & 1999 SCMR 2744 & Federation of Pakistan vs Muhammad Tariq Pirzada & 5
	\\
    \hline
    \end{tabular}
    \caption{List of the ten most cited SCMR judgements.}
    \label{tab:scmr}
\end{table*}

\subsection{Most cited Articles of the Constitution}
The Articles of the Constitution which are referenced in a judgement/order can
also be used as an indicator of the general scope of the case. For instance, if
any judgement refers to Article 6, which deals with treason, it might indicate
that the case revolves around treason. We again used a series of regular
expressions to extract this information. In total, 822 Articles and Sub-articles
of the Constitution are referenced in these judgements/orders. The ten most
cited Articles are listed in Table~\ref{tab:articles}. Article 184 (3), which
gives the Court its Suo Moto powers, is referenced in a fourth of all
judgements. Articles that deal with fundamental rights of citizens are pervasive
in the most cited list. In addition, two of the referenced Articles from the
Introductory Chapter in the table, in essence, also attend to fundamental
rights. Specifically, Article 3 enunciates the ``elimination of exploitation''
while Article 4 revolves around the ``right of individuals to be dealt with in
accordance with law''. Overall the ubiquity of these Articles in the judgements
suggests that the Court predominantly dealt with issues related to the
fundamental rights of the citizens. This in part might be due to its application
of judicial activism.

\subsection{Most cited Precedents/Cases}
Under the rules of the \emph{Common Law}, rulings are based on legal precedents
rather than any actual laws (in contrast to \emph{Statuary
Law})~\cite{holmes2009common}. Therefore, the set of previous cases that a
judgement cites--in essence, its legal lineage--can illustrate the general theme
of the judgement. For instance, if any judgement in Pakistan references
``Federation of Pakistan and others vs.
Moulvi Tamizuddin Khan'', it means that the judgement in some way revolves
around the ``Doctrine of State Necessity''. This case has been used since the
1950s to justify military rule in the country. Generally, previous judgements
are referenced by citing the law journal in which they were subsequently
published.
The most comprehensive of these journals, which contains all judgements since
independence for all courts, is \emph{Pakistan Law Decisions} (PLD). The
citation format is: \emph{PLD year-of-the-judgement
abbreviation-of-the-judgement-court judgement-number}. For instance, the above
mentioned ``Federation of Pakistan and others vs. Moulvi Tamizuddin Khan'' case
is cited as \emph{PLD 1955 FC 240}. Another important journal is \emph{Supreme
Court Monthly Review} (SCMR) which contains more recent cases heard in the
Supreme Court. Its citation format is: \emph{year-of-the-judgement SCMR
judgement-number}. In addition to these two, a number of other journals also
exist but we omit their details and analysis because they are rarely cited in
our dataset. To extract PLD and SCMR citations, we constructed regular
expressions to match their citation formats. Surprisingly, these simple
expressions were extremely effective in combing the documents for the required
information: \emph{L7 -- Even if the data is unstructured and noisy,
domain-specific filtering is very effective}. A total of 363 unique PLD
references and 910 SCMR references were obtained.
Table~\ref{tab:pld} and Table~\ref{tab:scmr} enumerate the 10 most cited PLD
judgements and SCMR judgements, respectively. Most of these cases deal with
constitutional matters such as the legality of martial law (PLD 1972 SC 139) and
the scope of ``public importance'' (1998 SCMR 793). In addition, some cases deal
with corruption (2012 SCMR 773) and the importance of the Civil Service (2010
SCMR 1301).

\section{Conclusion and Future Work}\label{sec:conclusion}
We presented an analysis of more than decade worth of open data from the Supreme
Court of Pakistan. Our analysis shows that post-restoration the judiciary in
Pakistan has undertaken cases of wider public interest that primarily protect
the fundamental rights of the citizens. In the process we highlighted several
challenges that any practitioner who makes use of open data will face,
including missing data, misspellings, and lack of term consistency.

As this was an initial study, our future work is extensive. This paper only
presented statistical analysis to understand some of the inherent trends in the
data. Our future work consists of using machine learning and data mining
techniques to classify the documents. For instance, leveraging the underpinnings
of common law, it would be interesting to train a learner to predict the outcome
of a case based on its similarity with prior cases. Furthermore, some of the key
concepts within the text can be understood better by linking in their semantic
meaning. Therefore, we intend on using DBpedia\footnote{DBpedia --
\url{http://dbpedia.org/}} to inject such semantics into the text. For example, if a case mentions an individual or a
company, ontological and descriptive information can be pulled in to discern the
nuances of legal text. Finally, we hope to extend our analysis to a wider range
of open data from the developing world; not necessarily confined to the legal
domain.

{\footnotesize \bibliographystyle{acm}
\bibliography{humanities}}

\end{document}